\documentclass[preprint,aps,superscriptaddress,floatfix]{revtex4}
\usepackage{graphicx}
\usepackage{dcolumn}
\usepackage{bm}
\usepackage{amsthm}
\usepackage{graphicx,subfigure}
\usepackage{epsfig}
\usepackage{float,placeins}
\usepackage{scalerel}
\usepackage[utf8]{inputenc}
\usepackage{physics}
\usepackage{xcolor, soul}
\sethlcolor{yellow}

\begin{document}

\preprint{APS/123-QED}

\title{Parametric Autoresonance with Time-Delayed Control}

\author{Somnath Roy}
\email{roysomnath63@gmail.com}
\affiliation{Department of Applied Mechanics and Biomedical Engineering,Indian Institute of Technology Madras,\\ Chennai,Tamilnadu,600036,India}
\author{Mattia Coccolo}%
\email{mattiatommaso.coccolo@urjc.es}
\affiliation{Nonlinear Dynamics, Chaos and Complex Systems Group, Departamento de Física, Universidad Rey Juan Carlos, Tulipán
s/n, 28933 Móstoles, Madrid, Spain
}
\author{Miguel A.F. Sanjuán}%
\email{miguel.sanjuan@urjc.es}
\affiliation{Nonlinear Dynamics, Chaos and Complex Systems Group, Departamento de Física, Universidad Rey Juan Carlos, Tulipán
s/n, 28933 Móstoles, Madrid, Spain }

\begin{abstract}

We investigate how a constant time delay influences a parametric autoresonant system. This is a nonlinear system driven by a parametrically chirped force with a negative delay-feedback that maintains adiabatic phase locking with the driving frequency.  This phase locking results in a continuous amplitude growth, regardless of parameter changes. Our study reveals a critical threshold for delay strength; above this threshold, autoresonance is sustained, while below it, autoresonance diminishes. We examine the interplay between time delay and autoresonance stability, using multi-scale perturbation methods to derive analytical results, which are corroborated by numerical simulations. Ultimately, the goal is to understand and control autoresonance stability through the time-delay parameters.
\end{abstract}

\maketitle

\section{Introduction}
We consider a specific type of nonlinear system in which the system becomes phase-locked with a driving force in an adiabatic manner, where the frequency of the forcing gradually changes over time. This phenomenon, widely known as autoresonance, leverages the system's nonlinearity to maintain resonance and effectively increase the system's amplitude over an extended period, despite changes in the system's parameters \cite{sanjuanbook,fajans2001autoresonant}. Typically, when the parameters of a nonlinear system change, the resonant frequency also shifts, causing the system to become detuned from resonance. In autoresonance, however, the system remains naturally phase-locked with the driving force in an adiabatic manner, allowing for self-adjustment of both amplitude and frequency. Since the initial observation of autoresonance \cite{ms_livingstone}, a plethora of studies have been conducted across various fields. 

The use and application of autoresonance have been explored in atomic physics \cite{meerson1990strong,liu1995nonlinear}, plasma physics \cite{fajans1999autoresonant1,fajans1999autoresonant2,andresen2011autoresonant,baker2015electron}, nonlinear wave interactions \cite{friedland1998autoresonance3,friedland1998autoresonant4,yaakobi2013complete}, planetary dynamics \cite{malhotra1999migrating,friedland2001migration,lanza2022tidal}, and fluid dynamics \cite{friedland1999control}. 

The theoretical framework of autoresonance has been extensively developed over the past few years. Some of the foundational work has been carried out by L. Friedland et al. \cite{friedland1997variational,fajans2001dampingeffect,fajans1999collective,fajans2000secondharmonic,nakar1999passage,barth2014quantum}. Studies can be found on asymptotic analysis and stability of autoresonance by L.A. Kalyakin et al.  \cite{kalyakin2008asymptotic,kalyakin2013stability}. Kovaleva et al. \cite{kovaleva2013limiting} have contributed an array of  works, including the limiting phase trajectory description of autoresonance  and investigations of autoresonance in nonlinear coupled chains \cite{manevitch2016autoresonant,kovaleva2018autoresonance,kovaleva2016autoresonance}. Additionally, R. Chac{\'o}n et al. \cite{chacon2005energy,chacon2010universal,chacon2008breakdown} have developed an energy-based theory and conducted a research on chaos and the breakdown of autoresonance . While autoresonance induced by external forcing has garnered significant attention, relatively few studies have focused on parametric autoresonance \cite{khain2001parametric,assaf2005parametric,kiselev2007capture,friedland2016parametric}.

Here, we have investigated the effect of time delay on a parametric autoresonant system. The control of resonance response whether deterministic or stochastic, in various nonlinear systems with time delay has proven to be an efficient strategy over the years \cite{hu1998resonances,mei2009effects,jeevarathinam2011theory,cantisan2020delay,zakharova2017time,maccari2003vibration}. In a recent work, the effect of delay has been discussed in a externally driven autoresonant array of Duffing-Ueda oscillator \cite{chacon2024}. Although time delay has been employed in systems with externally applied chirped forcing, its role in parametric autoresonant systems warrants attention due to its potential for effective applications across diverse fields. Chirped forcing refers to time-dependent parametric excitation where the driving frequency varies gradually over time. This gradual change allows phase-locking between the system and the driving force, a mechanism central to sustaining autoresonance despite parameter shifts. While autoresonance driven by external forces has been extensively studied, understanding the dynamics of parametric forcing with time-delayed feedback offers new opportunities for controlling resonance stability and amplitude growth.

In this work, we aim to analyze how time-delay feedback influences the stability of parametric autoresonance. Then, to derive a critical threshold for delay strength that governs the transition from decaying to growing oscillations. finally we validate these analytical predictions using numerical simulations. The ability to tune resonance behavior through time delay broadens potential control strategies in mechanical, electrical, and fluid systems. Driven by an interest in parametric autoresonance, we aim to investigate the effect of time delay in such systems, an aspect that has received limited attention in the existing literature, to the best of our knowledge. Our findings demonstrate that the delay strength can serve as a mechanism to control the growth of the autoresonant system.\\
The structure of this article is as follows. Section $\mathrm{II}$ outlines the essential mathematical formulations. In Sect. $\mathrm{III}$, we present numerical results to validate our analytical findings. Lastly, Sect. $\mathrm{IV}$ provides the concluding remarks.

\section{The model}
The mathematical description of a parametric autoresonant model with negative delayed feedback is given by
\begin{equation}
\ddot{x}+\gamma\dot{x}+\omega_0^2(1+h\cos\nu)x-\alpha x^3+kx(t-\Bar{\beta})=0,
\label{eq1}
\end{equation}
where the parametric forcing $h\cos\nu$, where the frequency now depends on time and is defined as $\dot{\nu}=2\omega-2\mu t$.The term $\nu(t)$ in the model evolves as specified before with $\mu$ representing the chirp rate, enabling phase-locking and continuous resonance. This follows from the definition of chirped forcing that refers to time-dependent parametric excitation where the driving frequency varies gradually. Here $\gamma$ is the damping constant, and $\omega_0,\alpha$ are the coefficients of linear and nonlinear stiffness, respectively. The model incorporates a constant delay $\bar{\beta}$ with strength $k$. Now introducing a dimensionless time $\omega t=\tau$, we can rewrite Eq.~(\ref{eq1}) as
\begin{equation}
    x''+\Gamma x'+\frac{\omega_0^2}{\omega^2}(1+h\cos(2\tau-\zeta\tau^2))x-\Lambda x^3+gx(\tau-\beta)=0,
    \label{eq2}
\end{equation}
where the prime ($'$) denotes the derivative with respect to $\tau$. The newly introduced constants are defined as $\Gamma=\gamma/\omega$, $\Lambda=\alpha/\omega^2$, $g=k/\omega^2$, $\beta=\omega\bar{\beta}$ and $\zeta=\mu/\omega^2$.

The equation of the model (Eq.~\ref{eq1}) models a parametric autoresonant system with delayed feedback, a generalization of Duffing-Ueda-like systems to incorporate time-delay effects. Time delay is critical for practical implementations where control and feedback loops introduce lag due to sensing and actuation dynamics. Real-world analogs include mechanical systems like pendulums with modulated pivots or beams with variable stiffness, as well as LC circuits where capacitance or inductance varies over time. In these contexts, sensors and controllers facilitate parametric forcing and delay management.

\subsection{Slow-flow and amplitude equation}
It has been observed that a critical value of the delay exists, above which autoresonance occurs, leading to a continuous growth in energy. Below this critical delay, however, the system's amplitude decays. To achieve this, the method of multiple scales has been applied to the Eq.~(\ref{eq2}) in order to derive the slow-flow dynamics of the system. Here, we consider all the parameters $h,\Gamma,\Lambda,g$ to be of the order $\mathcal{O}(\epsilon)$, and introducing a detuning parameter $\tilde{\sigma}$, such that $\omega=\omega_0+\epsilon\tilde{\sigma}$, gives $\omega_0^2/\omega^2\sim 1-\epsilon\sigma$, where $\sigma=2\omega_0\tilde{\sigma}/\omega^2$. Now expanding \(x\) as a power series
\begin{equation}
    x(\tau_0,\tau_1)=x_0(\tau_0,\tau_1)+\epsilon x_1(\tau_0,\tau_1)+\mathcal{O}(\epsilon^2)+..
    \label{eq3}
\end{equation}

We assume that \(x\) depends on two-time scales: a fast time scale \(\tau_0\) and a slow time scale \(\tau_1\). More generally, these time scales and their corresponding derivatives are defined as follows
\begin{equation}
\begin{split}
\tau_n&=\epsilon^n\tau;\\
\frac{\rm d}{\rm d\tau}&=\frac{\partial}{\partial\tau_0}+\epsilon\frac{\partial}{\partial\tau_1}\equiv D_0+\epsilon D_1,\\
\frac{\rm d^2}{\rm d\tau^2}&=\frac{\partial^2}{\partial\tau_0^2}+2\epsilon\frac{\partial^2}{\partial\tau_1 \partial\tau_0}+\epsilon^2\frac{\partial^2}{\partial\tau_1^2}\\
&\equiv D_0^2+2\epsilon D_0D_1+\epsilon^2D_1^2.
\end{split}
\label{eq4}
\end{equation}

Now substituting Eq.~(\ref{eq3}) and (\ref{eq4}) into Eq.~(\ref{eq2}) and separating the values by the order of $\epsilon$, we get the zeroth order $\mathcal{O}(\epsilon^0)$ equation as

\begin{equation}
    D_0^2x_0+x_0=0
    \label{eq5}
\end{equation} 
and $\mathcal{O}(\epsilon^1)$ equation as:

\begin{equation}
    D_0^2x_1+x_1=-2D_0D_1x_0-\Gamma D_0x_0-h\cos(2\tau_0-\zeta\tau_0^2)x_0+\Lambda x_0^3-gx_0(\tau_0-\beta)+\sigma x_0.
    \label{eq6}
\end{equation}

Now from Eq.~(\ref{eq5}) the zeroth order solution reads
\begin{equation}
    x_0(\tau_0,\tau_1)=A(\tau_1)\cos(\tau_0+\eta(\tau_1)),
    \label{eq7}
\end{equation}
where the amplitude $A$ and the phase $\eta$ now depend on the slow time $\tau_1$. Substituting Eq.~(\ref{eq7}) into Eq.~(\ref{eq6}) yields,

\begin{equation}
\begin{split}
D_0^2x_1+x_1=&2D_1A \sin(\tau_0+\eta)+2A\cos(\tau_0+\eta)D_1\eta+\Gamma A\sin(\tau_0+\eta)\\
&-\frac{hA}{2}\cos(2\psi)\cos(\tau_0+\eta)-\frac{hA}{2}\sin(2\psi)\sin(\tau_0+\eta)+\frac{3\Lambda A^3}{4}\cos(\tau_0+\eta)\\
&-gA\cos\beta\cos(\tau_0+\eta)-gA\sin\beta\sin(\tau_0+\eta)+\sigma A\cos(\tau_0+\eta)
\label{eq8}
\end{split}
\end{equation}

The newly introduced time-varying phase difference $\psi$ in Eq.~(\ref{eq8}) is given by
\begin{equation}
    \psi=\eta+\frac{\zeta\tau_0^2}{2}.
    \label{eq9}
\end{equation}

The coefficients of the $\sin(\tau_0+\eta)$ and $\cos(\tau_0+\eta)$ terms lead to secular terms in the solution for $x_1$. By setting these terms to zero, the resulting flow equation is governed by the following system of equations

\begin{equation}
    D_1A=-\frac{\Gamma A}{2}+\frac{hA}{4}\sin2\psi+\frac{gA}{2}\sin\beta=F_1(A,\psi)
    \label{eq10}
\end{equation}
\begin{equation}
    D_1{\psi}=\zeta\tau_0-\frac{3\Lambda A^2}{8}-\frac{\sigma}{2}+\frac{h}{4}\cos2\psi+\frac{g}{2}\cos\beta=F_2(A,\psi).
    \label{eq11}
\end{equation}

\begin{figure}[h!]
\begin{center}
\includegraphics[width=16.0cm,height=6.5cm]{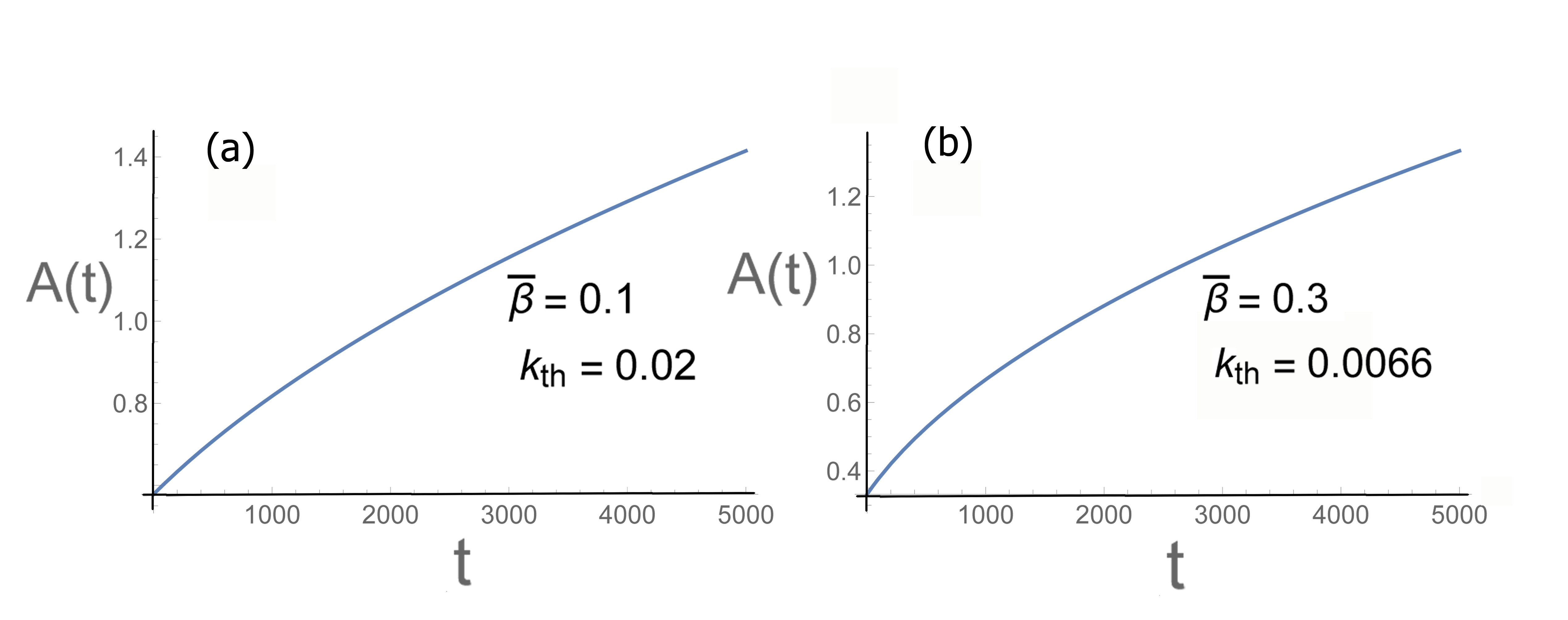}
\caption{Analytical plot of amplitude growth over time using Eq. (13) for parameters \(\gamma = 0.002\), \(\omega = 0.5\), and \(\alpha = 0.08\). Threshold values: (a) \(k_{\text{th}} = 0.02\) for \(\bar{\beta} = 0.1\) and (b) \(k_{\text{th}} = 0.0066\) for \(\bar{\beta} = 0.3\) indicate the onset of autoresonant growth when crossed.}
\label{fig:0}
\end{center}
\end{figure}

Now, consider a nontrivial quasi-steady fixed point ($\hat{A},\hat{\psi}$) for which $D_1A=D_1\psi=0$. Applying this condition to Eqs.~(\ref{eq10}--\ref{eq11}), followed by squaring and summing the results, yields an analytical expression for the detuning parameter $\sigma$

\begin{equation}
\sigma=\frac{1}{\omega^2}\bigg[\bigg(2\mu t+k\cos(\omega\bar{\beta})-\frac{3\alpha \hat{A}^2}{4}\bigg)\pm\bigg(\omega^4h^2-4(\omega\gamma-k\sin(\omega\bar{\beta}))^2\bigg)^{1/2}\bigg].
\label{eq12}
\end{equation}
Using the previously defined relation $\omega_0^2/\omega^2\sim 1-\epsilon\sigma$, we substitute the expression for $\sigma$ to arrive at the analytical expression for the quasi-steady amplitude $\hat{A}$, given as

\begin{equation}
 \hat{A}=\bigg[\frac{4}{3\alpha}\bigg((2\mu t+k\cos(\omega\bar{\beta})\pm\sqrt{\omega^4h^2-4(\omega\gamma-k\sin(\omega\bar{\beta}))^2}-\frac{\omega^2-\omega_0^2}{\epsilon} \bigg)\bigg]^{1/2}.
 \label{eq13}
\end{equation}
Equation (\ref{eq13}) shows the amplitude growth over time, as shown for the parameter values $\gamma=0.002$, $\omega=\omega_0=0.5$, $\sigma=0.1$, $h=0.0002$ and $\alpha=0.08$, in Fig.~\ref{fig:0}. In the following section, we analyze the stability of the system, showing that for stable, autoresonant amplitude growth, the delay strength must exceed a specific threshold. Below this threshold, the fixed points become unstable, making analytical expressions infeasible; however, numerical analysis can still be performed.

\subsection{Stability and the critical delay strength}

Next, we proceed to assess the stability of the system by determining the linearized Jacobian matrix at the fixed points given as

\begin{equation}
   J\equiv \begin{bmatrix}
\frac{\partial F_1}{\partial A} & \frac{\partial F_1}{\partial \psi}\\
\frac{\partial F_2}{\partial A} & \frac{\partial F_2}{\partial \psi}
    \end{bmatrix}_{(\hat{A},\hat{\psi})}.
    \label{eq14}
\end{equation}

The corresponding characteristic equation can be expressed as

\begin{equation}
    \det \begin{bmatrix}
-\frac{\Gamma}{2}+\frac{h}{4}\sin2\hat{\psi}+\frac{g}{2}\sin\beta-s ~~~~& \frac{h\hat{A}}{2}\cos2\hat{\psi}\\
-\frac{3\Lambda \hat{A}}{4} ~~~~& -\frac{h}{2}\sin2\hat{\psi}-s
    \end{bmatrix}=0
\end{equation}
using equations Eq.~(\ref{eq10}--\ref{eq11}), this can be further simplified to

\begin{equation}
    s^2+\Tr(J)s+\det(J)=0,
    \label{eq16}
\end{equation}

where
\begin{equation}
\begin{split}
    \Tr(J)=&-\frac{\Gamma}{2}-\frac{h}{4}\sin2\hat{\psi}+\frac{g}{2}\sin\beta\\
    \det(J)=&\frac{h}{2}\sin2\hat{\psi}\big(\frac{\Gamma}{2}-\frac{h}{4}\sin2\hat{\psi}-\frac{g}{2}\sin\beta\big)+\frac{3}{8}\Lambda h \hat{A}^2\cos2\hat{\psi}.
    \end{split}
\end{equation}

The steady-state vibration is asymptotically stable if, and only if, both of the following inequalities are simultaneously satisfied $\Tr(J)>0$ and $\det(J)>0$. Now, recognizing that 
$D_1A$ and $D_1\psi$ vanish at the fixed point $(\hat{A},\hat{\psi})$, we obtain the critical threshold for the delay strength, $k_{th}$, from the condition $\Tr(J)=0$ and by using Eq.~(\ref{eq10}). This provides a control mechanism for the parametric autoresonant system, in contrast to the conventional control, which has typically been achieved through the chirp rate $\mu$. The theoretical threshold value of $k_{th}$ can be expressed as
\begin{equation}\label{eq:kth}
    k_{th}=\frac{\gamma\omega}{\sin(\omega\bar{\beta})}.
\end{equation}
 For $\Bar{\beta}=0.1$, $\gamma=0.002$, and $\omega=0.5$ the threshold value is $k_{th}=0.02$ and for $\Bar{\beta}=0.3$, the threshold is $k_{th}=0.0066$ according to Eq.~(\ref{eq:kth}). When these thresholds are exceeded, autoresonant growth is observed. In the subsequent section, we provide numerical demonstrations showing that autoresonant behavior is triggered when this critical threshold is crossed and diminishes below it. In the following section, we numerically demonstrate how autoresonant behavior is triggered above this critical threshold and diminishes below it.

\section{Numerical results}

We validate the analytical predictions by conducting numerical simulations across a range of system parameters. Specifically, we examine the amplitude evolution and phase-space dynamics under varying delay strength $k$ and time-delay $\Bar{\beta}$ values. We quantify the autoresonance onset by fitting the oscillation envelope using an exponential model $x = ae^{bt}$.  We used an exponential fit to analyze the behavior of the oscillation amplitude and its envelope. The parameter \(b\) in the fit indicates different growth regimes: \(b > 0\) signifies increasing amplitude, \(b < 0\) represents decreasing amplitude, and \(b \approx 0\) denotes constant amplitude growth. This approach effectively distinguishes between growing, decaying, and steady-state oscillations. In our simulations, we have fixed $\gamma=0.002,\mu=0.00001,\omega=0.5,\alpha=0.03$ and we have varied $h,k,$ and $\Bar{\beta}$.  An additional study examining the impact of varying the initial conditions on the phenomenon was conducted, with no qualitative differences in the outcomes observed.

\begin{figure}[htbp]
  \centering
   \includegraphics[width=15.0cm,clip=true]{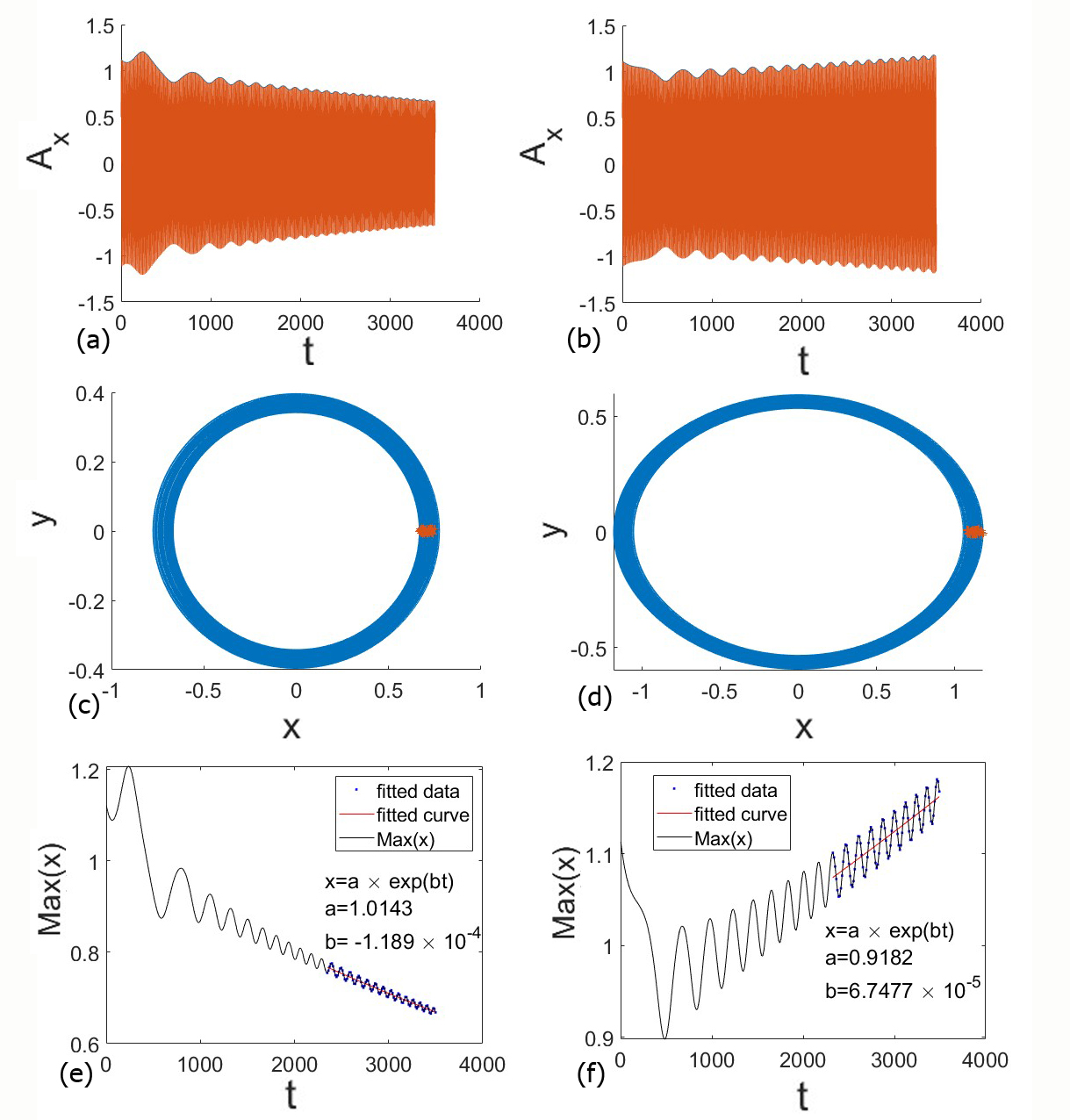}
   \caption{Panels (a) and (b) show the oscillations of the system with their maximum envelope. Panels (c) and (d) show the phase space orbits. Panels (e) and (f) show the maximum of the oscillations and the fit curve of the asymptotic part. The parameters are $h=0.002,~\Bar{\beta}=0.0092$, and  $k=0.02$ in the panels in the left column, while in the panels in the right column  $k=0.024$. In panels (e) and (f) we show the maximum envelope of the oscillations with the fitted curve and its equation.}
\label{fig:1}
\end{figure}

Contrarily to the theoretical analysis, shown in Fig.~\ref{fig:0}(a), for the case where $k \leq 0.02$, the numerical results indicate a negative trend in the maximum amplitude of the oscillations. In fact, numerically the threshold value of $k_{th}=0.0201$, even if it is a little bit higher, it is still in good agreement with the theoretical analysis.  In Fig.~\ref{fig:1}(a), Fig.~\ref{fig:1}(c), and Fig.~\ref{fig:1}(e) this is evident both visually in the decay of the oscillation envelope and mathematically in the negative sign of the exponential parameter $b$. Conversely, when $k = 0.024$, the behavior changes;  in Fig.~\ref{fig:1}(b), Fig.~\ref{fig:1}(d), and Fig.~\ref{fig:1}(f) the oscillations exhibit a growing trend, confirmed graphically and by a positive $b$, which signifies that autoresonance has been initiated for this parameter value. Worth to be mentioned is that a similar study has been conducted for the case of Fig.~\ref{fig:0}(b). Also, for $\Bar{\beta}=0.3$ the numerical threshold value of the parameter $k$ is similar to the value found from Eq.~\ref{eq:kth}, being the numerical value $k_{th}\approx0.0068$. These numerical findings are in strong agreement with the predictions of the threshold value of $k$ made by our analytical results, obtained from Eq.~\ref{eq:kth}.
 
\begin{figure}[htbp]
  \centering
   \includegraphics[width=14.0cm,clip=true]{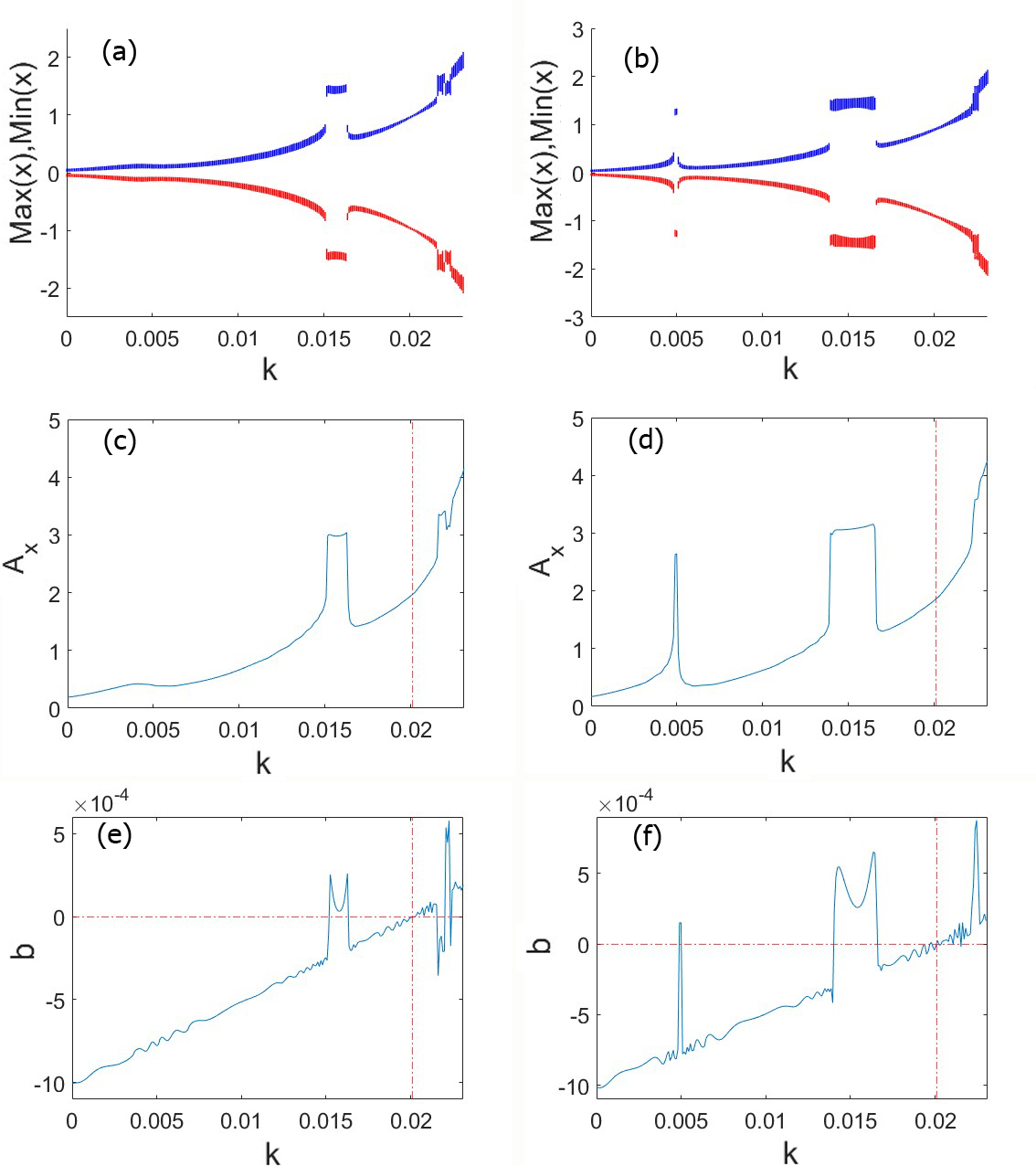}
   \caption{The figure shows the maximum and minimum diagram, the oscillations amplitude, and the exponent $b$ of the fitting of the oscillations maximum versus the parameter $k$, for $\Bar{\beta}=0.1$ and $h=0.001$ in the panels in the left column and for $h=0.002$ in the panels in the right column. As expected from the analytical analysis the autoresonance starts,  in both cases, for $k = 0.0201 \simeq k_{th}$, which is indicated by the vertical red lines in panels (c)-(f). The horizontal red lines in figures (e) and (f) indicate $b=0$, which marks the point where the value of $b$ changes from negative to positive, signifying the onset of the autoresonance.
 }
\label{fig:2}
\end{figure}

Now, we study the dependence of the autoresonance on the parameter $k$ for $h = 0.001$, $h = 0.002$, and $\Bar{\beta}=0.1$ as shown in Fig.~\ref{fig:2}. We observe that the trends of the curves for the oscillations amplitude and the exponent $b$ are similar, and in both cases, the value $k = 0.0201 \simeq k_{th}$ marks the definitive growing of the oscillations amplitude and the sign change of $b$. The numerical value of the threshold $k$ shows good agreement with the theoretical prediction $k_{th}=0.02$. Interestingly, for $h = 0.001$, there is a small negative peak in the $b$ curve after the threshold value of $k$, which is not present in the case where $h = 0.002$. On the other hand, in both cases, the oscillation amplitude grows after the threshold value of $k$. It is important to highlight that there is a set of $k$ values, around $k \approx 0.015$, and the value $k\approx 0.005$ in the $h=0.002$ case where autoresonance begins before the analytically predicted threshold values. This suggests a more complex interaction among the parameters to trigger the phenomenon. Nevertheless, our analytical results are confirmed by the findings presented in the figure. 
 Since we have done a similar study for $0.001\leq h \leq 0.002$ and the curves for the oscillation amplitude and the exponent $b$ have always similar trends with minor differences in some scattered peak, we now fix $h = 0.002$ for the remainder of our numerical analysis without loss of generality.

\begin{figure}[htbp]
  \centering
   \includegraphics[width=15.0cm,clip=true]{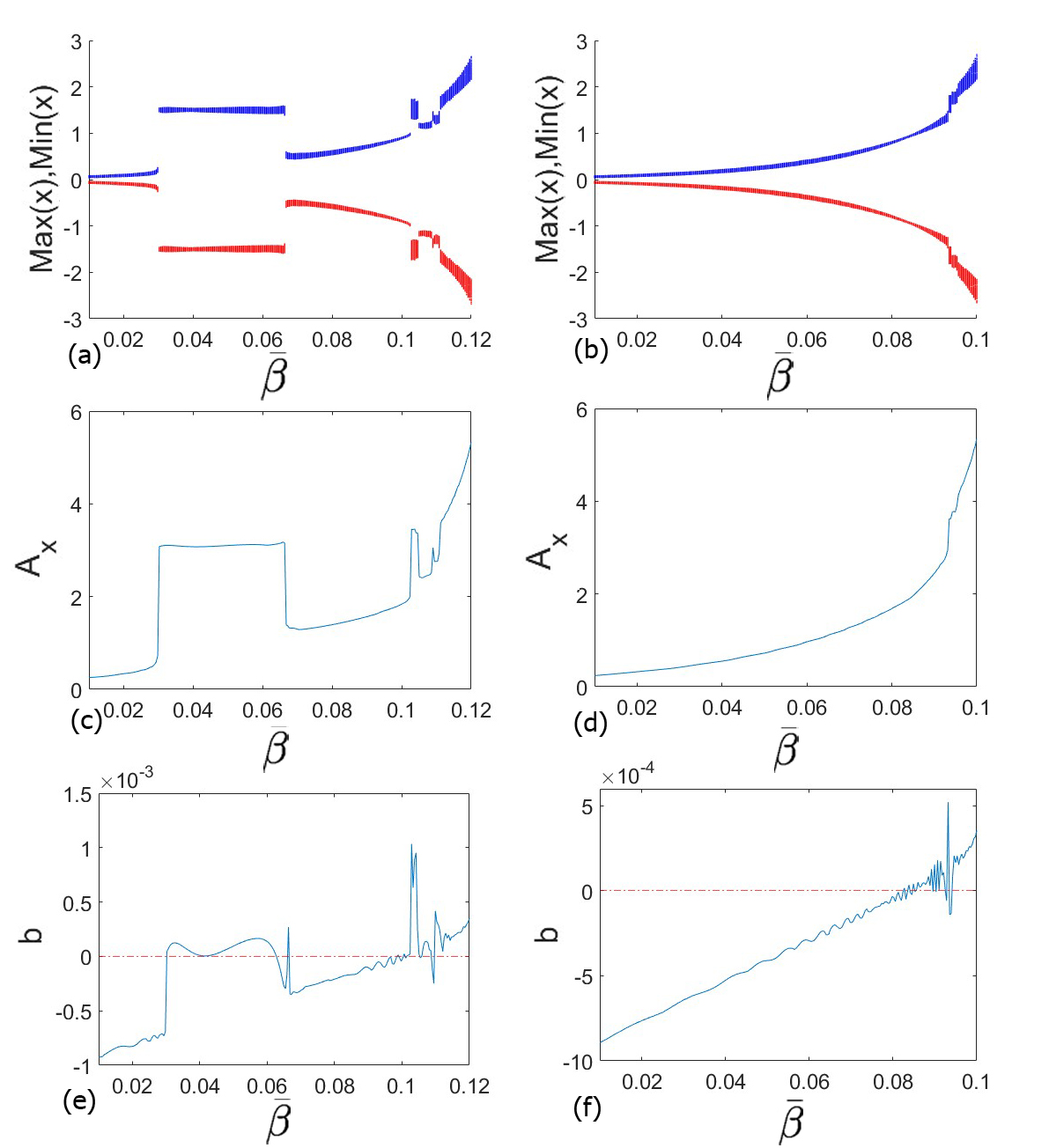}
   \caption{Figures (a) and (b) show the maximum and minimum diagram, figures (c) and (d) the oscillations amplitude, and figures (e) and (f)  the exponent $b$ of the exponential fitting curve of the oscillations maximum versus the parameter $\Bar{\beta}$, for $h=0.002$. We have fixed $k=0.02$ for the panels in the left column and $k=0.024$ for the panels in the right column. The horizontal red lines mark the zero to underline the change from negative to positive of the $b$ value, which indicates that the autoresonance is taking place. }
\label{fig:3}
\end{figure}

\begin{figure}[htbp]
  \centering
   \includegraphics[width=10.0cm,clip=true]{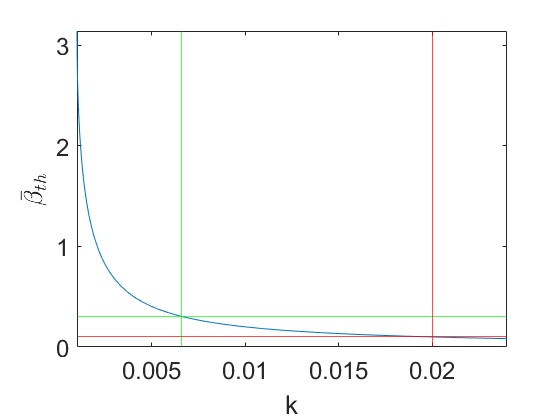}
   \caption{The figure shows the theoretical dependence of $\Bar{\beta}_{th}$ as a function of $k$ as resolved in Eq~\ref{eq:taut}. The red and green lines illustrate the cases already exposed in Fig.~\ref{fig:0}, $\Bar{\beta}=0.1$ and $k=0.02$ and $\Bar{\beta}=0.3$ and $k=0.0066$, respectively.}
\label{fig:3a}
\end{figure}

\begin{figure}[htbp]
  \centering
   \includegraphics[width=15.0cm,clip=true]{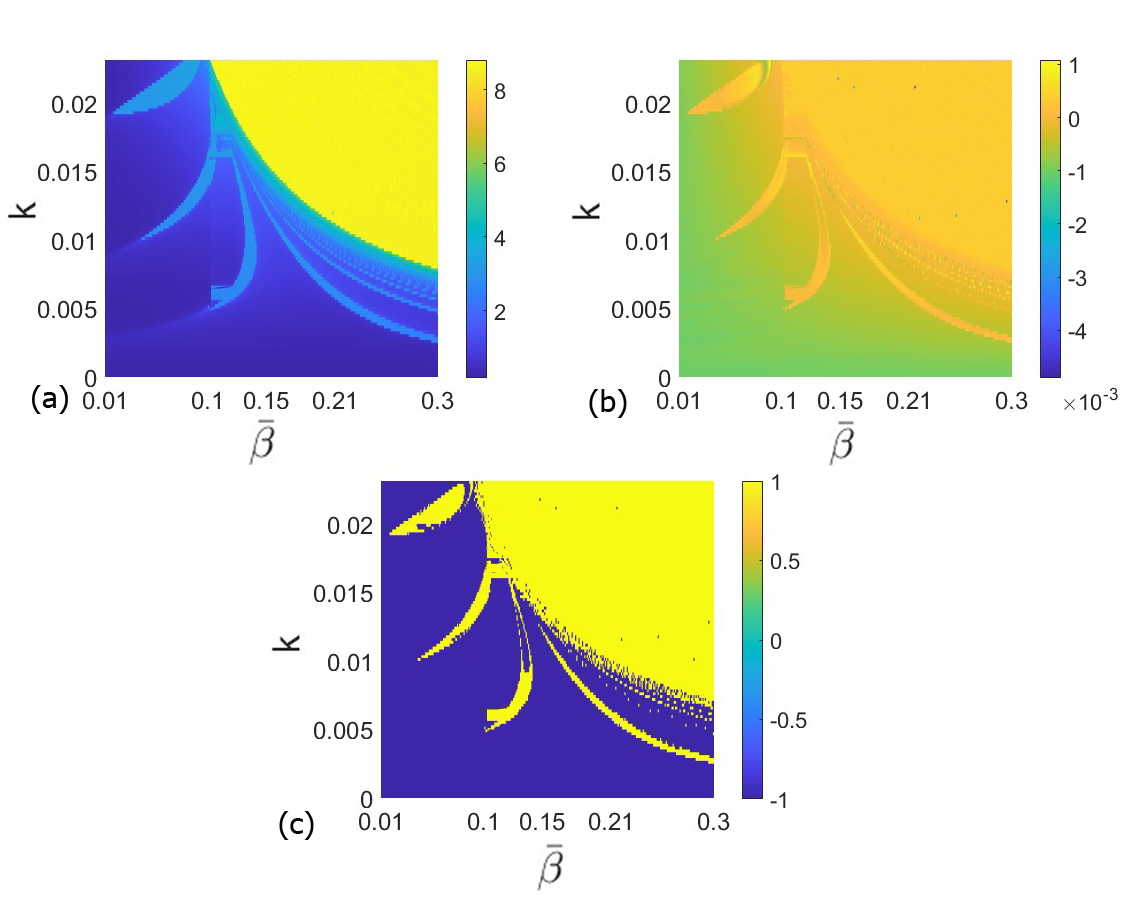}
   \caption{ The figure shows the gradient plot in the $k$--$\Bar{\beta}$ space for $h=0.002$ of the amplitude of the oscillations in panel (a), of the $b$ value of the exponential fit in panel (b), and the sign of the $b$ value in panel (c).  In this last figure, the yellow areas represent where $b$ becomes positive, meaning the $k$ and $\Bar{\beta}$ values related to these zones give birth to the autoresonance phenomenon. On the other hand, the blue ones represent negative values of the fit parameter, meaning the $k$ and $\Bar{\beta}$ values related to these zones do not trigger the phenomenon.  }
\label{fig:4}
\end{figure}

As the previous analytical analysis have shown, another key parameter that plays a significant role in triggering the autoresonance and is fundamental to our model is the time delay $\Bar{\beta}$. Therefore, in Fig.~\ref{fig:3}, we fix $h = 0.002$ and repeat the same numerical analysis as in Fig.~\ref{fig:2}, but now as a function of the parameter $\Bar{\beta}$. In this case, we study the maxima and minima diagram, the oscillation amplitude, and the behavior of the exponent $b$ for $k = 0.02$ (left panels) and $k = 0.024$ (right panels). The figure highlights distinct behaviors for the two cases, although they follow a similar trend: autoresonance is triggered as values of $\Bar{\beta}$ increase. Specifically, the threshold value of $\Bar{\beta}$ is higher for the case where $k = 0.02$. This outcome aligns with the expectations derived from analyzing Eq.~\ref{eq:kth}. In fact, we can rewrite it as
\begin{equation}\label{eq:taut}
    \Bar{\beta}_{th} =\frac{1}{\omega}\arcsin{ \left(\frac{\gamma\omega}{k}\right)},
\end{equation}
and this equation suggests an inverse relationship between $k$ and $\Bar{\beta}$, as illustrated in Fig.~\ref{fig:3a}. Additionally, for $k = 0.024$, the oscillation amplitude follows an almost monotonic increasing trend, except near the threshold value of $\Bar{\beta}$. In contrast, the $k = 0.02$ case exhibits high oscillation amplitudes, as shown in Fig.~\ref{fig:3}(c) and Fig.~\ref{fig:3}(f), for $0.03 < \Bar{\beta} < 0.07$, which result from autoresonance triggered before the threshold value of $\bar{\beta}$, as confirmed by Fig.~\ref{fig:3}(a), Fig.~\ref{fig:3}(c) and Fig.~\ref{fig:3}(e). Thus, we conclude that for smaller values of $k$, larger values of $\Bar{\beta}$ are required to definitively initiate autoresonance, and vice versa, as predicted in Fig.~\ref{fig:0}, but for smaller values of $k$ the interaction between the two parameter is more complex than theoretically forecasted. Additionally, the amplitude of the oscillations grows with increasing $\Bar{\beta}$ once exceeds the threshold value.  

To conclude and for completeness, we vary both parameters, $\Bar{\beta}$ and $k$, to visualize the dependence of oscillation amplitude and the exponent $ b $ on these two parameters, as shown in Fig.~\ref{fig:4}. In this figure, stripes and a yellow area in the top-right region of the panels indicate the onset of autoresonance. Specifically, in Fig.~\ref{fig:4}(a), these stripes and the yellow area highlight regions of maximum oscillation amplitude. The boundary of the yellow region in the top-right marks threshold values of $ k $ and $ \Bar{\beta} $ where autoresonance is definitively established. For values within this region, the oscillation amplitude continues to grow. For accuracy, it should be noted that when measuring the oscillation amplitude within this yellow area, the integration—and thus the amplitude calculation—was stopped at $ y = 3 $ to prevent divergence. In Fig.~\ref{fig:4}(b), the stripes indicate values for the exponent $ b $, and in Fig.~\ref{fig:4}(c), they show regions where $ b $ is positive. These plots reveal two important insights. Firstly, the yellow area in the top-right confirms the theoretical trend observed earlier, indicating that for smaller $ k $, the value of $ \Bar{\beta} $ required to trigger autoresonance increases. Secondly, the stripes suggest that the interaction between the two parameters may be more complex than the theoretical predictions.

\FloatBarrier
\section{Conclusions}

In conclusion, this study highlights the significant role of time delay in parametric autoresonance, a topic that has received limited attention in the existing literature. Through analytical formulations and numerical simulations, we have shown that the delay strength can serve as a critical control mechanism for the growth of the autoresonant system. In the numerical analysis, we confirmed the findings from our theoretical study. Specifically, we observed that the numerical threshold value $k = 0.0201$, at which autoresonance begins, is in good agreement with the theoretical prediction and it is independent of the parameter $h$, as expected from the analytical study.  To demonstrate this, we have plotted the maxima-minima diagram, the oscillation amplitude, and the exponent $b$ of the fitting exponential curve, which characterizes the asymptotic behavior of the oscillator. The exponent $ b $ is negative when the asymptotic behavior decreases and positive when it increases, indicating the onset of autoresonance. Using these plots, we also explored the dependence of the phenomenon on the time delay $\Bar{\beta}$. Our results show that for smaller values of $ k $, the resonance initiates at larger values of the time delay. Again, the numerical results corroborate the theoretical prediction. Finally, we illustrated the dependence of autoresonance on both $ k $ and $\Bar{\beta}$ with different gradient plots. Here, we can see that although the theoretical prediction of the trend of $\Bar{\beta}$ as a function of $k$ is respected, this relation can be more complex. Our findings contribute to a deeper understanding of the dynamics involved in parametric autoresonance and pave the way for potential applications across various fields, such as atomic physics, plasma dynamics, and engineering systems.

In summary, we have demonstrated that the presence of time delay significantly influences the stability and growth of a parametric autoresonant system. Using both analytical techniques and numerical simulations, we have identified a critical delay strength \(k_{\text{th}}\) that marks the onset of sustained amplitude growth. Our numerical simulations confirm that for values of \(k > k_{\text{th}}\), the oscillation amplitude increases, while for \(k < k_{\text{th}}\), it decays, aligning well with the theoretical predictions derived from Eq.~(18). Additionally, we found that the interplay between \(k\) and the delay time \(\bar{\beta}\) exhibits a more complex relationship than initially predicted, highlighting areas for future exploration in the control of parametric resonance.

These findings pave the way for practical applications where precise resonance control is needed, such as mechanical systems with variable stiffness or feedback-controlled electronic circuits, to further broaden the understanding of delay-induced resonance phenomena.

\section{Acknowledgment}

This work was supported by the Spanish State Research Agency
(AEI) and the European Regional Development Fund (ERDF, EU)
under Project Nos. PID2019-105554GB-I00 and PID2023-148160NB-I00 (MCIN/AEI/10.13039/
501100011033).

\end{document}